\documentclass[article,reqno]{amsart}
\usepackage{amsthm}
\usepackage{amsfonts}
\usepackage{amsmath}
\usepackage{amssymb}
\usepackage{mathtools}
\usepackage{latexsym}
\usepackage{epsfig}
\usepackage{graphicx}
\usepackage{booktabs}
\usepackage{color}
\usepackage{graphicx}
\usepackage{tikz}
\usepackage{array,booktabs}
\usepackage{physics}

\newcommand{\beq}{\begin{equation}}
\newcommand{\eeq}{\end{equation}}
\newcommand{\beqnl}{\begin{eqnarray}}
\newcommand{\eeqnl}{\end{eqnarray}}

\newcommand{\be}{\begin{equation}}
\newcommand{\ee}{\end{equation}}
\newcommand{\bes}{\begin{equation*}}
\newcommand{\ees}{\end{equation*}}
\newcolumntype{C}{>{\centering\arraybackslash}m{1in}} 

\numberwithin{equation}{section}

\title{Analogue Gravity: the Subcritical Case}
\author{F. Belgiorno$^{1,2}$,
S.L. Cacciatori$^{3,4}$, 
and  A. Vigan\`o$^{4}$}
\address{\noindent $^1$Dipartimento di Matematica, Politecnico di Milano, Piazza Leonardo 32, IT-20133 Milano, Italy\endgraf
$^2$INdAM-GNFM, Piazzale Aldo Moro 5, Roma, Italy.\endgraf
$^3$Dipartimento di Scienza e Alta Tecnologia, Universit\`a dell'Insubria, Via Valleggio 11, IT-22100 Como, Italy\endgraf
$^4$INFN sezione di Milano, via Celoria 16, IT-20133 Milano Italy\endgraf
}

\begin{document}

\begin{abstract}

We focus on the subcritical case in Analogue Gravity for a subluminal dispersion relation, and investigate the problem of pair creation in the absence of an horizon analytically, for the subcritical white hole case. We limit our attention to weakly dispersive media. 
In particular, we take into account for simplicity the Corley model and adopt an approach  
inspired by quantum mechanics in the so-called overbarrier scattering. The presence of complex turning points is shown to play a fundamental role in allowing a mode conversion which would be absent in a naive WKB approach. We  show that there is still a pair creation phenomenon, which, in the subcritical case, is non-Planckian and without  a Hawking partner. Our fully analytical results are qualitatively in agreement with the literature at hand. Furthermore we confirm 
the existence of a critical regime betwen the subcritical and the transcritical ones where no pair creation occurs.

\end{abstract}

\maketitle

\section{Introduction}

We study the problem of the so-called subcritical case for the Hawking effect in Analogue Gravity. 
The term subcritical indicates the absence of a real horizon in the medium one is taking into account, 
and this phenomenon has been extensively studied in the physical literature, because 
of the Vancouver experiment with water \cite{weinfurtner-prl,weinfurtner-book}, which involves a white hole situation.  
The actual absence of a horizon has been found in subsequent analysis \cite{Parentani_numerical,euve,parentani-sub}.  It is also worth mentioning that the first experiment with water was discussed in \cite{rousseaux-original}, and that both the subcritical and the transcritical cases for water were studied also in \cite{euve-noise,euve-cocurrent,fourdinoy-weakly}. \\
Even if several analytical calculations have been devoted to the transcritical case, see e.g the following non exaustive list of references ~\cite{brout,corley,himemoto,saida,Schutzhold-Unruh,balbinot,unruh-s,cpf,Leonhardt-Robertson,Coutant-prd,Coutant-und,Un-schu,petev,Coutant-thick,hopfield-hawking,linder,philbin-exact,hopfield-kerr,coutant-sub,coutant-kdv,coutant-bdg,Konig,master,bec,s-matrix,trevisan,trevisan-exact,analog}, the subcritical case has been analytically treated just in a few papers. The first study is the one contained in  \cite{coutant-subcritical}, where the Bremmer expansion \cite{bremmer49,bremmer} (see also \cite{nakano-third}), adapted to an ordinary differential equation 
of the $n$-th order, has been provided and exploited for an analytical study of the subcritical case.  Subsequent analytical calculations, involving also the subcritical case for water and BEC, again in the limit of low dispersive effects, are found in \cite{coutant-kdv,coutant-bdg}, and exploit a KdV equation emerging in the approximation of no co-propagating modes \cite{coutant-kdv}. It is also worthwhile to mention that tunneling method has been applied in \cite{delporro} both in the transcritical and in the subcritical case.\\
For the case of the analogue Hawking effect in dielectric media, horizonless situations were numerically taken into account in \cite{fincaru,finazzi-carusotto}
and analytical studies appear in \cite{trevisan}, where a perturbative study for the dielectric case appears, and also in \cite{trevisan-exact}, 
where an exact solution for a suitable smooth refractive index profile is provided, and analytical results, both in the transcritical case and in the subcritical one, 
appear.\\
In the present paper, we exploit a weak dispersion expansion of the fourth order differential equation governing the phenomenon, where
$\epsilon$ is the suitable weak expansion parameter. In  \cite{master,bec,s-matrix,analog}, for the transcritical case we have shown that an equation of the Orr-Sommerfeld type inherited by Nishimoto's works \cite{ni,ni-I,ni-global,ni-tp} enables us to treat in a quite unified 
way the transcritical case for black hole configurations\footnote{White hole configurations are also covered by the aforementioned framework.}, which amounts to the presence of a real turning point (real horizon). 
In the subcritical case, there is no real turning point. Still, we show that in our equation complex turning points may appear, and this is fundamental in order to 
allow particle creation even in the situation where no real horizon is present. 
In order to make as clear as possible our picture, we shall discuss the most simple case, represented by the Corley model \cite{corley}, discussed also for the 
transcritical case in \cite{cpf,master,analog}. With the same aim, we shall also provide, in a short interlude in standard quantum mechanics, an example in order to 
show how a complex turning point for the WKB approximation plays a fundamental role in allowing the presence of a reflected mode in the case where 
the so-called overbarrier scattering occurs and, according to a naive WKB approach, no reflection would appear.\\
Then, we discuss the subcritical case for a would-be white hole situation, i.e. for a smooth step-like velocity profile for the Corley model, which, on the grounds of \cite{master,bec,analog}  can be 
extended straightforwardly to water, dielectric media and also to the subcritical black hole case.\\
We show that: (a) amplitudes for the outgoing dispersive modes $P,N$ can be calculated analytically, and the mode conversion process which gives rise to the dispersive modes is potentially heavily suppressed with respect to the transcritical case; 
(b) the behaviour of the number of created $P,N$ modes, as the frequency $\omega\to 0$, is proportional to $\omega$, as deduced also in numerical and in analytical 
studies \cite{Parentani_numerical,coutant-subcritical,trevisan}; (c) the spectrum is non-Planckian but it may be associated with an effective temperature; (d) the phenomenon shares some basic characteristics with the analog Hawking effect, even if, {\sl stricto sensu}, is not such; (e) a suitable parameterization of the flow velocity $v(x)$ allows to identify a critical value of the parameter such that, at the transition between subcritical and transcritical regimes no mode conversion occurs. The reason is that, at the transition between the two regimes, the potential turning point is shifted to infinity.\\
The plan of the paper is the following. In sec. \ref{orsosec} we resume some fundamental information about the formalism developed in \cite{master,bec,s-matrix,analog}. 
In sec. \ref{corley-subcritical}, we apply the aforementioned formalism to the Corley model in the subluminal case, and in absence of a real turning point (real horizon). We discuss also the differences in the scattering between the transcritical case and the subcritical one. In sec. \ref{overb-landau}, in order to prepare the ground for the analysis which is developed in the subsequent section \ref{scattering}, we present the problem of the overbarrier scattering in quantum mechanics. In sec. \ref{scattering}, we fully develop the analysis of the Hawking-like scattering process in the subcritical case, and in sec. \ref{critical} we extend our considerations, with the aim of confirming, at least qualitatively, that there is room for a possibly efficient pair creation also in the subcritical case, and discuss the limit case between the subcritical and the transcritical regimes. In sec. \ref{concl} we summarize our results and provide a further discussion. Two appendices are also provided in support of our analysis in the main text.

\section{The master equation: a Orr--Sommerfeld type fourth order equation}
\label{orsosec}

As we have shown in \cite{master,bec}, three significant cases of wave equations in dispersive analogue gravity can be 
related to the equation
\beq
\label{orso}
\epsilon^2 \frac{d^4 \Phi}{dx^4}\pm\biggl[ p_3 (x,\epsilon) \frac{d^2 \Phi}{dx^2}+
p_2 (x,\epsilon) \frac{d \Phi}{dx} +p_1 (x,\epsilon) \Phi \biggr]=0, 
\eeq
where the upper sign occurs in the case of subluminal dispersion and the lower one in the case of superluminal dispersion.
The latter case is considered in Nishimoto's works (see e.g.~\cite{ni} and references therein). Furthermore, 
\beq
p_i (x,\epsilon) = \sum_{n=0}^\infty p_{i n} (x) \epsilon^n ,
\eeq
is assumed.
As $\epsilon \to 0$ one finds the so-called reduced equation
\beq
p_{30} (x) \frac{d^2 \Phi}{dx^2}+
p_{20} (x) \frac{d \Phi}{dx} +p_{10} (x) \Phi =0 .
\eeq
Solutions of 
\beq
p_{30}(x)=0
\eeq
define the turning points (TPs) of the equation, and in the analysis of the reduced equation the behaviour of solutions in the neighbourhood of the TPs is of utmost relevance for the 
scattering problem we mean to delve into, even in the case of a complex turning point (CTP). 
In~\cite{ni} it is assumed that the reduced 
equation displays a Fuchsian singularity at the TP (no matter if real or complex).\\ 

We recall also that near the TP $x=a$, both in the real and in the complex case  \cite{ni}, the original equation is replaced by the fourth-order differential equation 
\beq
\label{four-z}
\frac{d^4 w}{dz^4}\pm \biggl( z \frac{d^2 w}{dz^2} +\lambda \frac{dw}{dz} \biggr)=0,
\eeq
where, as discussed in \cite{master}, $w(z)$ represents the (rescaled \cite{master}) wave-function in the so-called near horizon approximation, to be then matched in the so-called linear region, with WKB solutions which hold far from the turning point. We refer the reader to 
\cite{master} for further details. In \eqref{four-z} the upper sign is for the subluminal case and the lower one is for the superluminal one, and 
\beq
\label{change-var}
z =(p'_{30}(a))^{1/3} \epsilon^{-2/3} (x-a),
\eeq
and 
\beq\label{lambda-n}
\lambda = \frac{p_{20}(a) }{p'_{30} (a)}.
\eeq

\subsection{Near TP solutions and bases}
\label{neartp bases}

We sum up in the following our analysis in \cite{analog} for the subluminal case, and refer to \cite{analog} for the superluminal one.
We limit ourselves to notice that, with respect to the bases described in \cite{ni-global} for the superluminal case, the bases described below are obtained simply by a rotation $z\mapsto \exp (-i\pi/3) z$, so we have 
the replacement $\arg (z)\mapsto \arg (z) +\frac{\pi}{3}$. These bases are also confirmed by the analysis in \cite{langer1955}.
Solutions to equation~\eqref{four-z} are e.g. found, by adopting the method of the Laplace transform, as discussed in \cite{ni} for the superluminal case, by means of Laplace integrals
\beq
\label{gen-airy}
w_j(z) = \frac{1}{2\pi i} \int_{C_j} dt\ t^{\lambda -2} \exp (z t\pm \frac{1}{3} t^3),
\eeq
with a suitable choice for the paths $C_j$ in the complex $t$-plane. It must be stressed that such solutions are regular at the TP, and admit a series expansion there (see \cite{ni-global}). Of course, there is also a constant solution, which appears to be a Dirac delta in the space of the Laplace transform. This further solution is fundamental to obtain a complete basis for the scattering problem also in the region near the TP (i.e. near the horizon) \cite{master}. 
Solutions of equation \eqref{four-z} in the transcritical case were indexed by NH as near horizon solutions in \cite{master,analog}. 
The relevant paths for the bases are indicated in the following figures: 
\begin{figure}[hbtp!]
\begin{center}
\includegraphics[scale=0.45]{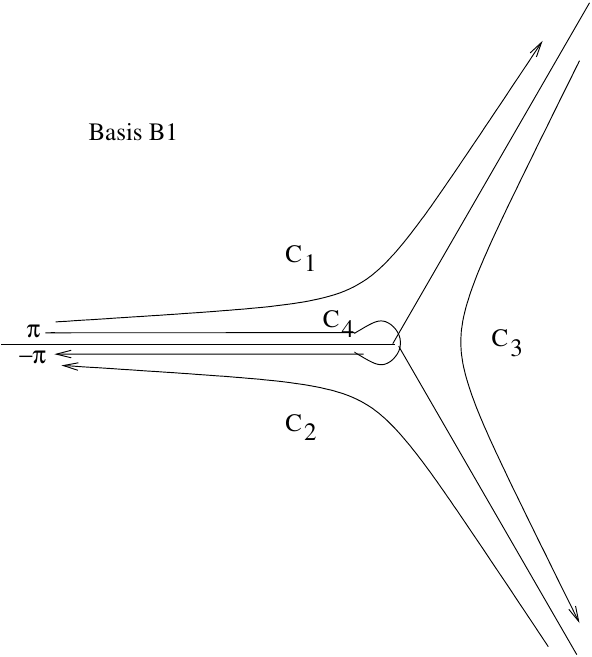} 
\end{center}
\caption{Paths labeled with $C_j$ in the $t$-plane, $j=1,2,3,4$ for the basis $[1,w_4(z),w_1(z),w_2(z)]$. There is a branch cut for $\arg (t)=\pi$.  }
\label{fig:paths-ni4}
\end{figure}
The first basis, say B1, which is relative to the sector $-\frac{2\pi}{3}<\arg (z) <\frac{2\pi}{3}$, is  $[1,w_4(z),w_1(z),w_2(z)]$. 
\begin{figure}[hbtp!]
\begin{center}
\includegraphics[scale=0.45]{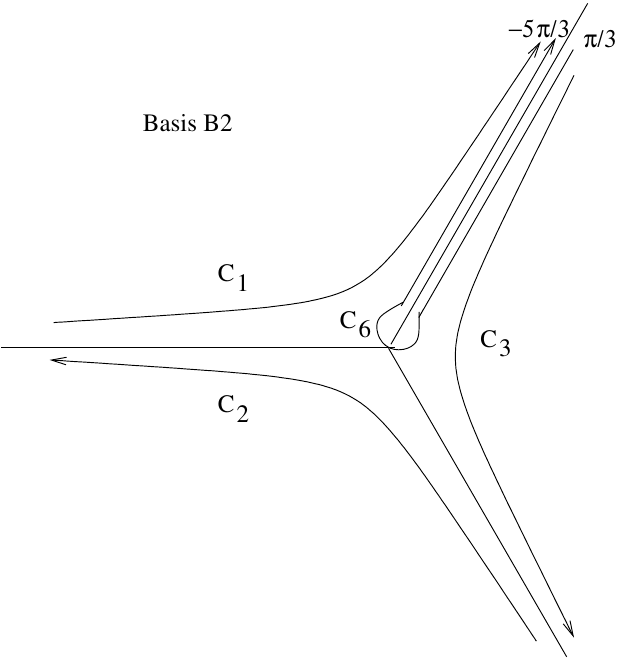} 
\end{center}
\caption{Paths labeled with $C_j$ in the $t$-plane, $j=1,2,3,6$ for the basis $[1,w_6(z),w_3(z),w_1(z)]$. There is a branch cut for $\arg (t)=\frac{\pi}{3}$.}
\label{fig:paths-ni6}
\end{figure}
The second basis, say B2, which is relative to the sector $0<\arg (z) <\frac{4\pi}{3}$, is $[1,w_6(z),w_3(z),w_1(z)]$. 
\begin{figure}[hbtp!]
\begin{center}
\includegraphics[scale=0.45]{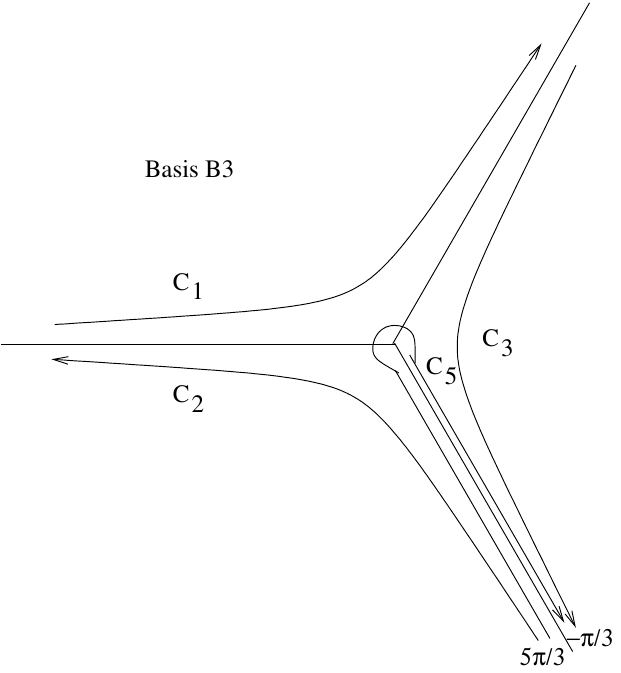} 
\end{center}
\caption{Paths labeled with $C_j$ in the $t$-plane, $j=1,2,3,5$ for the basis $[1,w_5(z),w_2(z),w_3(z)]$. There is a branch cut for $\arg (t)=\frac{5\pi}{3}$.}
\label{fig:paths-ni5}
\end{figure}
The third basis, say B3, which is relative to the sector $\frac{2\pi}{3}<\arg (z) <2\pi$, is $[1,w_5(z),w_2(z),w_3(z)]$.\\

\newpage

\section{An application to the simplest model: the Corley subluminal model}
\label{corley-subcritical}

We limit our considerations to the simplest dispersive model, which is the Corley model in the subluminal case \cite{corley-jacobson,corley}
\beq
S=\frac{1}{2} \int d^2x \Big[ \Big(\Big(\frac 1c \partial_t + \frac{v}{c} \partial_x\Big)\phi\Big)^2+\phi( \partial_x^2+ \epsilon^2 \partial_x^4) \phi\Big].
\eeq
The model is associated with a conserved inner product \cite{corley} and with current conservation. 
By separating variables as in \cite{corley}: $\phi (t,x)=e^{-i \omega t} \varphi (x)$, one obtains the fourth-order ordinary differential equation
\beq
\label{eqco}
\epsilon^2 \partial_x^4 \varphi + \biggl(1-\frac{v^2 (x)}{c^2}\biggr)  \partial_x^2 \varphi+ 2 \frac{v(x)}{c^2}  (i \omega -v'(x)) 
\partial_x \varphi- i \frac{\omega}{c^2} (i \omega -v' (x)) \varphi =0,
\eeq
where $v(x)$ is the velocity field and $v'(x)$ stands for its first derivative with respect to $x$, and $c$ is the constant sound velocity. 
The relevant scale is $\epsilon^2$, which is associated with the nonlinearity, as in~\cite{corley-jacobson,corley}.  By comparison 
with \eqref{orso}, we find
\beqnl
p_3 (x) &=& 1-\frac{v^2 (x)}{c^2},\\
p_2 (x) &=& 2 \frac{v(x)}{c^2}  (i \omega -v'(x)),\\
p_1 (x) &=& \frac{\omega}{c^2} (\omega +i v' (x)),
\eeqnl
and we can observe that the model can be baptized as ``the simplest'' because just no dependence on $\epsilon$ of the coefficients occurs. This is a peculiar characteristic of the Corley model, which instead does not occur in the 
case of dielectric media, BEC, water \cite{master,bec}. 
In the limit as $\epsilon\to 0$, one obtains the so-called reduced equation, 
\beq
\biggl(1-\frac{v^2 (x)}{c^2}\biggr)  \partial_x^2 \varphi+ 2 \frac{v(x)}{c^2}  (i \omega -v'(x)) 
\partial_x \varphi- i \frac{1}{c^2} \omega (i \omega -v' (x)) \varphi =0.
\label{reduced-co}
\eeq
For a complete and exhaustive calculation of both the WKB solutions and  the near-horizon solutions for the black hole transcritical case, we refer to \cite{master}.

\subsection{Subcritical vs. Transcritical: analogies and differences}

\begin{figure}[hbtp!]
\centering
\includegraphics[scale=0.5]{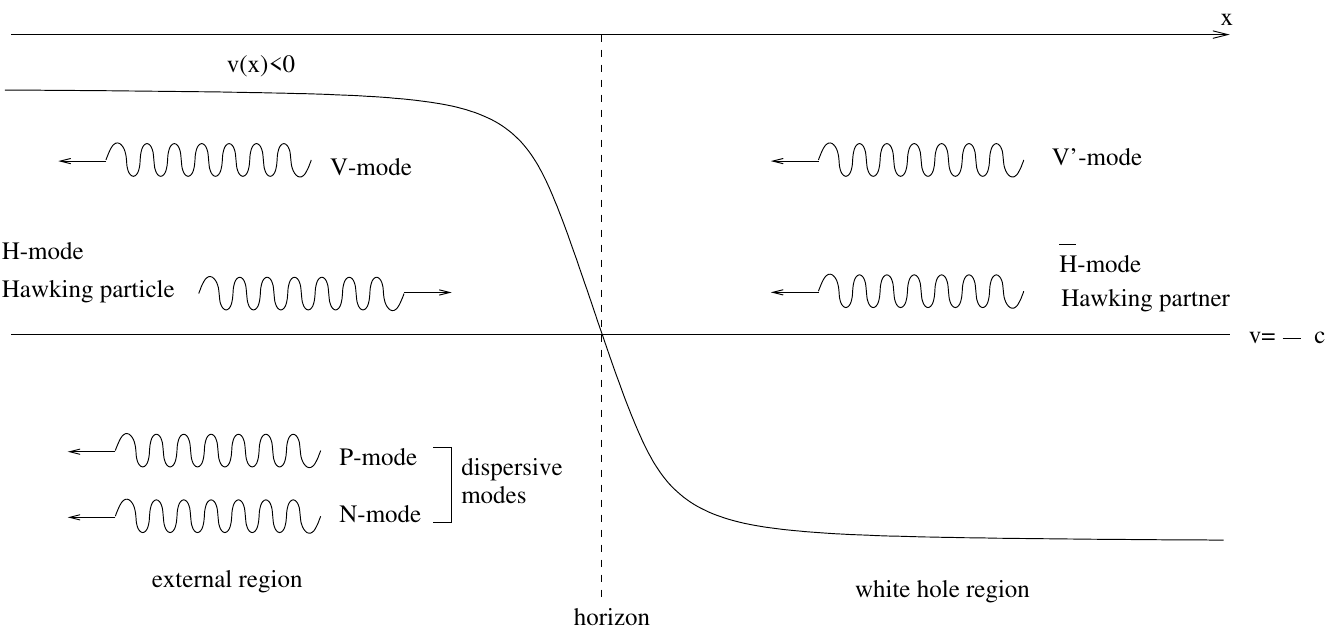} 
\caption{Propagating modes involved in the scattering in a white hole situation. The horizon corresponds to a real TP. We stress the presence of a Hawking partner, 
$\bar{H}$, which is an antiparticle state, as $N$ is. There are two further non propagating modes, 
the decaying mode $D$, and the growing mode $G$ \cite{cpf,corley,master}, the former playing a role in the Hawking process and the latter with zero amplitude. }
\label{fig-wh}
\end{figure}

\begin{figure}[hbtp!]
\centering
\includegraphics[scale=0.5]{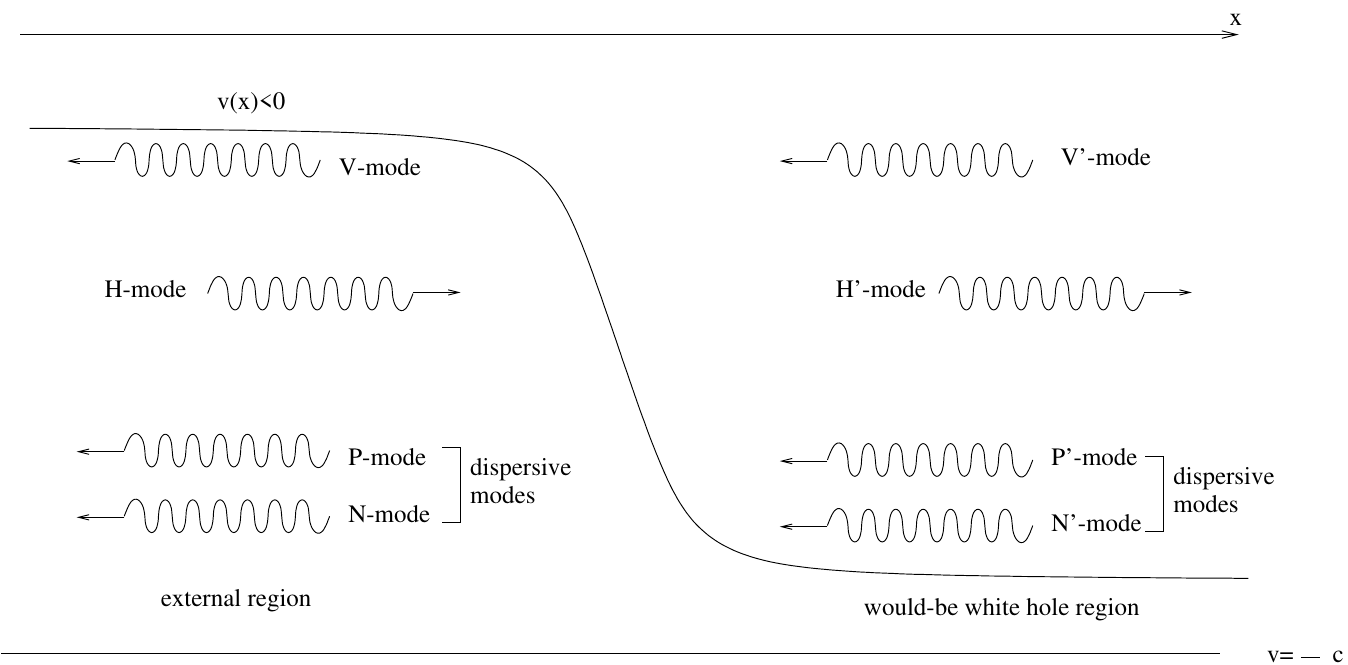} 
\caption{Propagating modes involved in the scattering in a would be white hole situation. There is no real horizon. The only antiparticle states appearing are 
$N,N'$. There is no Hawking partner, which is replaced by $H'$, which is a particle propagating to infinity.}
\label{fig-wh-sub}
\end{figure}

In the subcritical case, the absence of a horizon implies a number of very important distinct features of the 
scattering. 
In order to better understand the 
situation at hand, we refer to Fig. \ref{fig-wh}, which represents the propagating modes in presence of a real TP, i.e. of a white hole horizon, and to  Fig. \ref{fig-wh-sub}, 
where a real TP is instead missing, and then a subcritical case occurs. The modes represented in both the aforementioned figures correspond to the states occurring asymptotically, according to the 
dispersion relation at hand. They are also associated with WKB solutions of the equation of motion. We choose a step-like profile for the velocity field (see also  Fig. \ref{fig:v-sub-bh}). 
As it is evident, the latter situation involves two more propagating modes $P',N'$ which replace the decaying mode $D$ and the 
growing mode $G$ which occur in the transcritical case (but do not contribute to the $S$-matrix at infinity). 

In the subcritical case, the four modes in the external region remain unaltered. Still, in the would-be white hole region, drastical changes appear, indeed, apart for the 
spectator mode $V'$, we find three novel modes which are different from the ones in the black hole region of the transcritical case: two ingoing modes 
$P',N'$, moving towards the would-be horizon region from $x\to +\infty$ and one outgoing mode $H'$, moving towards $x\to +\infty$. $P',N'$ are dispersive modes 
($N'$ is an antiparticle). Cf. Fig. \ref{fig-wh-sub}.

\subsection{WKB solutions} 
\label{wkb-corley}

The following WKB solutions for the Corley model hold true \cite{master,analog}
\begin{align}
\varphi_{+} (x) & =  \Biggl( \frac{1}{1-\frac{v^2 (x)}{c^2}} \Biggr)^{3/4} \exp (  \frac{i}{\epsilon} \int_a^x ds \sqrt{1-\frac{v^2 (s)}{c^2}} ) 
\exp (  i \frac{\omega}{c} \int_a^x ds \frac{v (s)}{c} \frac{1}{1-\frac{v^2 (s)}{c^2}}), \label{p-mode}\\
\varphi_{-} (x) & =  \Biggl( \frac{1}{1-\frac{v^2 (x)}{c^2}} \Biggr)^{3/4} \exp (  -\frac{i}{\epsilon} \int_a^x ds \sqrt{1-\frac{v^2 (s)}{c^2}} ) 
\exp (  i \frac{\omega}{c} \int_a^x ds \frac{v (s)}{c} \frac{1}{1-\frac{v^2 (s)}{c^2}}) , \label{n-mode}
\end{align}
where \eqref{p-mode} represents the dispersive (non-normalized) positive norm modes $P,P'$ far from the TP,
and \eqref{n-mode} represents  the dispersive (non-normalized) 
negative norm ones $N,N'$. 
Near the TP $x=a$ we get 
\begin{align}
\varphi_{+} (x) & \simeq \biggl(\frac{2\kappa}{c} (x-a)\biggr)^{-3/4} (x-a)^{-\frac{i\omega}{2\kappa}}\exp \left(  \frac{i}{\epsilon} \frac{2}{3} 
\sqrt{\frac{2\kappa}{c}} (x-a)^{\frac{3}{2}}\right),\\
\varphi_{-} (x) & \simeq \biggl(\frac{2\kappa}{c} (x-a)\biggr)^{-3/4} (x-a)^{-\frac{i\omega}{2\kappa}}\exp \left( - \frac{i}{\epsilon} \frac{2}{3} 
\sqrt{\frac{2\kappa}{c}} (x-a)^{\frac{3}{2}}\right).
\end{align}

We stress that, as the above WKB solutions in a naive WKB approach hold everywhere on the real axis, nothing peculiar could be 
expected in such a naive approach, and, in particular, no non-trivial connection formulas for modes on the left and on the right 
in the asymptotic regions.\\
Two further solutions can be obtained from the reduced equation, and in general, unless exact solutions of the reduced equation can be explicitly found, as in 
\cite{analog}, 
 near the regular singular point $x=a$, which represents the CTP, we can provide the following series expansions (cf. \cite{ni}):
\begin{align}
\varphi_{+s} (x) &= (x-a)^{i \frac{\omega}{\kappa}}  \biggl( 1+\sum_{n=1}^{\infty} d_n (x-a)^n\biggr),\\ 
\varphi_{-s} (x) &= 1+\sum_{n=1}^{\infty} c_n (x-a)^n,\\
\end{align}
where  $\varphi_{+s} $ represents the asymptotic non-dispersive positive norm modes $H',H$ (would-be Hawking mode and Hawking partner), 
and $\varphi_{-s}$ represent further short wavenumber modes $V,V'$ which eventually play simply the role of a disconnected spectator mode, at least as far as the leading order process is concerned, just as in \cite{analog}.\\
It is useful to provide approximate solutions of the reduced equation even for large $x$ (in the external region with respect to the would-be black hole). It is easy to show that for large $x$ in the above sense we have $v(x)\sim$ const, and then $v'=0$.
As a consequence, e.g.~under the conditions of theorem 1.9.1 of~\cite{eastham}, we get as $x\to \infty$
\begin{align}
\varphi_{V} (x) &\sim \exp ( -i {\omega} \frac{1}{c-v_{r}} x),\\
\varphi_{H} (x) &\sim \exp ( i {\omega} \frac{1}{c+v_{r}} x).
\end{align}
with $\lim_{x\to \infty} v(x) =: v_r > -c$ and $v_r<0$. As $x\to -\infty$ we get the asymptotic behavior 
\begin{align}
\varphi_{V'} (x) &\sim \exp ( -i {\omega} \frac{1}{c-v_{l}} x),\\
\varphi_{H'} (x) &\sim \exp ( i {\omega} \frac{1}{c+v_{l}} x), 
\end{align}
with $\lim_{x\to -\infty} v(x) =: v_l > -c$ and $v_l<0$. Both these solutions have positive norm, so that $H'$ is not an antiparticle state, 
and is actually an ingoing state, i.e. a mode approaching $x=0$ from the left asymptotic region.\\ 
It is very remarkable that formulas inherited by \cite{corley} (WKB-like) actually correspond to {\sf exact solutions} of the reduced equation:
\begin{align}
\varphi^{ex}_{j} (x) &= \exp ( -i {\omega} \int^x dy \frac{1}{c-v (y)} ), \quad j=V,V' \label{vmode}\\
\varphi^{ex}_{k} (x) &= \exp ( i {\omega} \int^x dy \frac{1}{c+v (y)}),   \quad k=H,H' \label{hmode}
\end{align}
which still display the correct behavior both in the linear region and in the asymptotic one.

\begin{figure}[hbtp!]
\begin{center}
\includegraphics[scale=0.80]{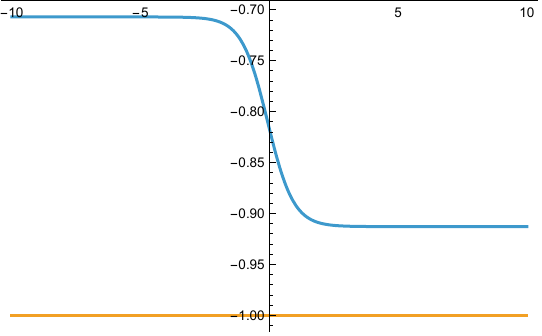} 
\end{center}
\caption{Qualitative plot for the velocity profile $v/c$, with $v$ as in \eqref{profv} and with $\gamma=1$. As seen, on the real axis the white hole horizon 
condition $v/c=-1$ is never reached.}
\label{fig:v-sub-bh}
\end{figure}

In the following, we shall assume the profile velocity 
\beq\label{profv}
v(x) = -c \sqrt{\frac{2}{3}+\frac{1}{6} \tanh \gamma x},
\eeq
where $\gamma$ is an inverse length scale, and the profile is a subcritical one. For a qualitative plot, see Fig. \ref{fig:v-sub-bh}. 
In the following, we shall discuss the case with $\gamma=1$ for simplicity. Further considerations about the case $\gamma\not = 1$ are found in sections \ref{critical} and \ref{concl}.

Turning points amount to solutions of 
\beq
p_{30}(x)=0 \longleftrightarrow x_{ctp}={\mathrm{arctanh}} (2)=:a,
\eeq
so that 
\beq
x^{(n)}_{ctp}=x_0+i \frac{\pi}{2} + i \pi n,
\eeq
where 
\beq
x_0=\frac{1}{2} \log (3),
\eeq 
and with $n\in \mathbf{Z}$,  and we select (cf. also \cite{coutant-sub} and see also sections \ref{overb-landau} and \ref{scattering})
\beq\label{cctp}
x_{ctp}=x^{(0)}_{ctp}=x_0+i \frac{\pi}{2}. 
\eeq
We also recall that the presence of a TP jeopardizes the WKB 
approximation, which cannot hold {\sl at} the TP. Still, it may be valid in the so called linear region \cite{cpf,master}, where both the WKB and the 
near TP approximation (cf. \eqref{four-z} and its solutions) hold true.\\
There is an unexpected property to be taken into account: as usual, one assumes that the surface gravity 
\beq
\kappa:=v'(x_{ctp})
\label{surf}
\eeq
represents the surface gravity of the white hole, which is proportional to the temperature. 
In the present case, one would expect that, as the velocity in  \eqref{profv} is decreasing, $\kappa$ would be negative. 
This is actually not the case, as far as a {\sf complex turning point} is involved. Indeed, what happens is that $\kappa=v'(x_{ctp})>0$, 
and this is not what is expected for a white hole, for which $\kappa<0$. For the profile \eqref{profv} with $\gamma=1$, one gets $\kappa=\frac{c}{4}$.

\subsection{Near TP solutions} 
\label{NTP-corley}

For completeness, we provide asymptotic expansions as $z\to \infty$ for the functions $w_i (z), i=1,2,3,4,5,6$, and recall that  in our case $\lambda=1-i\frac{\omega}{\kappa}$: 
\beqnl
w_1 (z) &\sim& \frac{1}{2 \sqrt{\pi}} e^{-\frac{3}{4} \pi i} e^{\frac{\pi \omega}{2 \kappa}} z^{-\frac{i \omega}{2 \kappa}-\frac{3}{4}} e^{\frac{2}{3} i z^{\frac{3}{2}}},\;  \quad \quad \quad B1,B2\\
w_2 (z) &\sim& \frac{1}{2 \sqrt{\pi}} e^{-\frac{5}{4} \pi i} e^{-\frac{\pi \omega}{2 \kappa}} z^{-\frac{i \omega}{2 \kappa}-\frac{3}{4}} e^{-\frac{2}{3} i z^{\frac{3}{2}}}, \quad \quad B1\\
w_2 (z) &\sim& \frac{1}{2 \sqrt{\pi}} e^{-\frac{11}{4} \pi i} e^{\frac{\pi \omega}{2 \kappa}} z^{-\frac{i \omega}{2 \kappa}-\frac{3}{4}} e^{\frac{2}{3} i z^{\frac{3}{2}}}, \quad \quad \quad B3 \label{w2a}\\
w_3 (z) &\sim& \frac{i}{2 \sqrt{\pi}} e^{-\frac{3}{4} \pi i} e^{-\frac{\pi \omega}{2 \kappa}} z^{-\frac{i \omega}{2 \kappa}-\frac{3}{4}} e^{-\frac{2}{3} i z^{\frac{3}{2}}}, \quad \quad B2,B3\\
w_4 (z) &\sim& \frac{i}{\pi} \Gamma \left(-\frac{\pi \omega}{\kappa} \right) \sinh \left( \frac{\pi \omega}{\kappa} \right) z^{\frac{i \omega}{\kappa}},\quad \quad \quad  \quad B1\\
w_5 (z) &\sim& \frac{i}{\pi} \Gamma \left(-\frac{\pi \omega}{\kappa} \right) \sinh \left( \frac{\pi \omega}{\kappa}\right) z^{\frac{i \omega}{\kappa}},\quad \quad \quad  \quad  B3\\
w_6 (z) &\sim& \frac{i}{\pi} \Gamma \left(-\frac{\pi \omega}{\kappa} \right) \sinh \left( \frac{\pi \omega}{\kappa} \right) z^{\frac{i \omega}{\kappa}}.\quad \quad \quad  \quad  B2
\eeqnl
We don't specify herein ranges of validity, as we mean that each expression holds true in the sectors associated with the bases described above and explicitely indicated on the right, except for two different expressions for $w_2(z)$ according to the basis chosen, in particular 
\eqref{w2a} must live on a different sheet.
Still, see also Appendix \ref{ba-nishi} and  Appendix \ref{ba-langer}, where the change of basis is described both from the point of view of \cite{ni,ni-global} but also from the point of view of \cite{langer1955}. 
When a change of basis occurs, lateral connection formulas \cite{ni,ni-global} arise, whose non trivial appearance is associated with the Stokes phenomenon;  together with central connection formulas, which are associated with the matching of the near turning point expression of a basis function with its WKB expression, central and lateral connection formulas give rise to connection formulas 
in the scattering process for a real TP. We refer to \cite{analog} for a full discussion. For the case of a CTP, see the discussion in sec. \ref{scattering}.\\
Before discussing the role of our CTP \eqref{cctp}, 
we discuss 
a simpler example in standard quantum mechanics.

\section{An interlude in standard Quantum Mechanics. The overbarrier scattering}\label{overb-landau}

\begin{figure}[hbtp!]
\begin{center}
\includegraphics[scale=0.55]{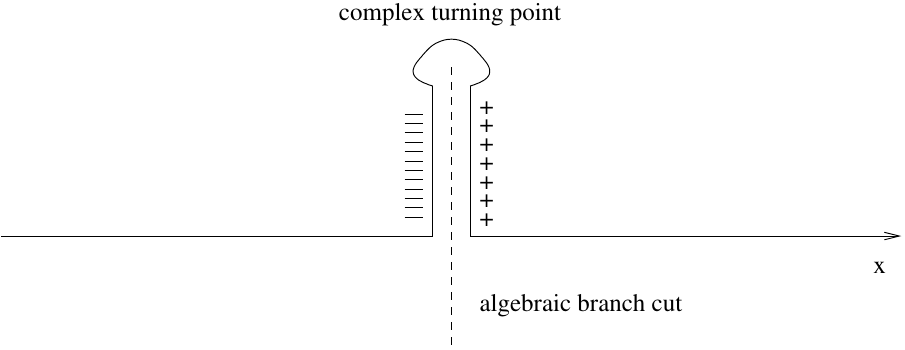} 
\end{center}
\caption{Path for turning around the complex turning point. The dashed line represents the algebraic branch cut starting from the complex turning point. }
\label{fig:landau}
\end{figure}

In this section, we deal with a case which is typical in quantum mechanics, and can be solved by appealing to the presence of CTP's. We shall solve it in agreement with 
the discussions appearing in \cite{landau} and reference therein, and also in \cite{fedoryuk}.\\ 
The problem is a standard, and is the following. Let us consider a one-dimensional problem, with a smooth, step-like potential energy $V(x)$ 
in the Schr\"{o}dinger equation, and with energy $E>\sup_{x} V(x)$. As a consequence, we fix that the energy is greater than the potential energy, in such a way 
that no real turning point is allowed. Let us consider an ingoing particle coming from the left with asymptotic momentum $k_->0$. 
In a semiclassical naive WKB approach, no reflection occurs for such a mode, which simply becomes an outgoing one reaching $x=+\infty$ with momentum $k_+>0$. 
Quantum mechanics allows the presence of a reflected mode, for the profile we propose in subsection \ref{wkb-bc-algebraic}, 
and also an exact solution for the reflection coefficient can be found in 
\cite{flugge}, problem 37 therein. How to become able to find a WKB solution where also a reflected mode appears? The answer to this question is a bit subtle, but 
is also well-known, and was found by Landau and developed further on by other Soviet scientists (see \cite{landau}, pp. 191-193). What emerges from the aforementioned studies 
is that one has to involve CTP's, which appear in our case as couples of complex conjugate turning points, and one has to select the ones that are 
as near as possible to the real axis. From a physical point of view, a complex turning point, as a real one, is a source of ``mode conversion'', in the specific sense 
that allows the appearance also of the reflected mode, which is instead missing in a naive WKB approach. From a mathematical point of view, near the 
CTP, one has to abandon the real axis, rise vertically up to the CTP, turn around it, and then come down again to the real axis, as in \cite{landau}. See Figure 
\ref{fig:landau}. This approach is also called phase integral method \cite{heading}. We stress that this procedure can be done just within the WKB approximation, under the hypothesis that the Landau path is still in the 
linear region when it passes very near the CTP.

As far as one deal with the scattering problem by using only WKB solutions, as in \cite{landau,fedoryuk,khriplovich}, this means that the CTP corresponds to an algebraic turning point, from which a vertical branch cut arises, and such that a negative momentum on the 
left of the branch cut appears, amounting to the otherwise missing reflected mode. Asymptotically to the left, such a negative momentum is equal to $-k_-$, as naively expected. 
Landau refers simply to a change of sign of the semiclassical momentum, as discussed in the following subsection. 
One must also consider that the WKB solutions cannot 
be too near the TP (where the approximation fails, as is well known).\\ 

In \cite{landau,fedoryuk,khriplovich},  the overbarrier scattering is studied by exploiting just WKB solutions and considering the Stokes phenomenon in the region where both approximations hold. 
See also \cite{pokro}. 

\subsection{WKB solutions, matching and the role of the algebraic branch cut}
\label{wkb-bc-algebraic}

As is well known, in the WKB approximation, the Schr\"{o}dinger equation admits solutions in the following form at the leading order in $\epsilon$:
\beq\label{wkb-qm}
\psi_{\pm}^{WKB} (x)=\frac{1}{q(x)^{1/4}} \exp \left(\frac{i}{\epsilon} \int_{a}^x \sqrt{q(x)} dx \right), 
\eeq
where $q(x):=E-V(x)$, with $V(x)$ the potential and we have set {\sf only in this section} $\epsilon^2:=\frac{\hbar^2}{2 m}$.\\
In the absence of real TP(s), there is no way to obtain a reflected mode if $E>U(x)$. Still, one can adopt a trick which was introduced by Landau, and is 
discussed in \cite{landau}. It is still true that the equation 
\beq
E=V(x)
\label{e-tp}
\eeq
may allow complex solutions which appear in couple of complex conjugates TP $x_\pm^{(m)}$, where $m$ labels the $m$-th couple. 
For simplicity, and for allowing a comparison with exact solutions, we assume that 
\beq\label{pot-flugge}
V(x)=\frac{V_0}{2}\left(1+\tanh \left(\frac{x}{2 a}\right)\right),
\eeq
as in \cite{flugge}, problem 37, pp. 86-89. For definiteness, we shall consider the case $V_0=2,a=\frac{1}{2}$, and, moreover, {\sf we set $E=3$ in \eqref{e-tp}, as this specific value will be useful for calculations in the white hole case.} 
Let us select, as in \cite{landau}, the couple nearest to the real axis, and choose the positive root, indicated as $x_+$ for simplicity. Of course, 
this root is complex and its imaginary part can be at finite distance from the real axis (in the sense that it is by no means true that necessarily it 
is at  a very small distance from the real axis even if $E$ is only slightly greater than $V(x)$). See below. 
It is easy to show that \eqref{e-tp} is satisfied for the complex value
\beq
x_{ctp}={\mathrm{arctanh}} (2),
\eeq
so that one obtains a sequence of CTP: 
\beq
x^{(n)}_{ctp}=x_0-i \frac{\pi}{2} + i \pi n,
\eeq
with $n\in \mathbf{Z}$. Let us also choose, as in \cite{landau}, $x_{ctp}=x^{(1)}_{ctp}=x_0+i \frac{\pi}{2}$. In this model, as evident, the imaginary part 
is at finite distance from the real axis.\\     
In the region around the turning point where the WKB approximation still holds, we obtain for the solutions in \eqref{wkb-qm}
\beq
\psi_{\pm}^{WKB} (x)\sim \frac{1}{\left(3(x-{\mathrm{arctanh}} (2))\right)^{1/4}} \exp \left(\pm\frac{i}{\epsilon} \frac{2 \sqrt{3}}{3} (x-{\mathrm{arctanh}} (2))^{3/2}\right), 
\eeq
We need to prescribe how our physical solutions are affected by the presence of such a CTP. 
Landau introduces a suitable path like the one in Fig. \ref{fig:landau}. 
What happens is that, for the square root $\sqrt{q(x)}$, the TP $x=a$ behaves like an algebraic branch cut, such that $\sqrt{q(x)}$ on the right of the 
branch becomes $-\sqrt{q(x)}$ on the left. This means that, if $\sqrt{q(x)}\to k_\pm$ as $x\to \pm \infty$, we gain in the left asymptotic region 
a reflected mode. Substantially, what happens is a mode-conversion phenomenon at the turning point, which plays an essential role in resorting 
the expected reflected mode which is missing in a naive approach to the WKB approximation for the overbarrier case. \\
In the Landau prescription, cf.  Fig. \ref{fig:landau}, which generates a shift in phases between the modes $\psi_\pm$ and the reflected mode $\psi_R$, 
a further factor $\exp (-\frac{\pi}{\epsilon})$ multiplies $\psi_R$ (and only $\psi_R$) whereas the amplitude of the incoming mode and the transmitted one are near the same: $|\psi_{tr}|\sim |\psi_{in}|$ \cite{landau,khriplovich}.\\ 
In the following, our strategy is similar to the one in  \cite{fedoryuk}.
Let us introduce the phase integral factor 
 \beq\label{phase}
S(x_{ctp},x):= \frac{1}{\epsilon} \int_{x_{ctp}}^x ds \sqrt{q(s)}=\frac{1}{\epsilon} \int_{x_{ctp}}^x ds \sqrt{(2-\tanh (s))} 
\eeq
and notice that the primitive associated is, also for complex $x$, 
\beq
G(x):=\mathrm{arctanh}(\sqrt{(2-\tanh (x))}-\sqrt{3}\; \mathrm{arctanh} \left( \frac{\sqrt{(2-\tanh (x))}}{\sqrt{3}} \right).
\eeq
Let us consider the ingoing mode, for which we need $S(x,x_{ctp})$, with $x\ll 0$ but finite. Starting just over the cut of  $\mathrm{arctanh}  z$ in the complex plane, given by 
$(-\infty,-1]\cup [1,\infty)$, one finds that the imaginary part  of $S$, which contributes to the amplitude of the mode, is ${\textrm{Im} S}=-\frac{\pi}{2 \epsilon}$, so that one obtain a real factor 
$F_{in}= \exp(\frac{\pi}{2 \epsilon})$.  We can obtain the same factor  for the transmitted mode, for which we need $S(x_{ctp},x)$, with $x\gg 0$ but finite, 
if we choose to approach the positive real axis cut from below. Then we find $F_{tr}= \exp(\frac{\pi}{2 \epsilon})$. This is the only way we can get 
the requirement of Landau for $|\psi_{tr}|\sim |\psi_{in}|$. The reflected mode requires $-S(x_{ctp},x)$, with $x\ll 0$ but finite, and the real factor we obtain is suppressed with respect to the aforementioned ones only if we approach the real negative $x$ from below the cut. Then we find $F_{R}= \exp(-\frac{\pi}{2 \epsilon})$.  
The final ratio in amplitude between the incoming mode, transmitted mode and reflected mode is given by 
\beq
\psi_{in} \mapsto \alpha \psi_{tr} + \beta \exp (-\frac{\pi}{\epsilon}) \psi_{R},
\eeq
where $\alpha,\beta$ are phase factors which are irrelevant for the calculation of the reflection and transmission coefficients.\\
For the reflection coefficient one finds \cite{landau,fedoryuk,khriplovich}
\beq
|R|^2=\exp (-\frac{2\pi}{\epsilon}).
\label{r-wkb}
\eeq

\subsection{A comparison with an exact solution}

In order to allow a better comparison, we consider exact solutions of a step-like smooth barrier in quantum mechanics (see \cite{flugge} for full details), 
and to understand a posteriori that the above approach performs very well, at least in the semiclassical limit, we compare the WKB result for the 
reflection probability, calculated as above, with the exact result provided by \cite{flugge} for the potential \eqref{pot-flugge}. 
We have the general result \cite{flugge}
\beq
|R|^2=\frac{\sinh^2 (\pi (\kappa-\sigma))}{\sinh^2 (\pi (\kappa+\sigma))},
\eeq
where, with our parameters, one finds  
\beq
\kappa\pm \sigma=\frac{\sqrt{E}}{2\epsilon}\pm \frac{\sqrt{E-2}}{2\epsilon}.
\eeq
Then, for $\epsilon \to 0$ and for $E=3$, we find that \eqref{r-wkb} holds true, and 
the agreement between the exact amplitude and the WKB one is actually very good.\\ 

We stress that the present discussion is not to be considered as novel, but is simply introductory to the trickier problem of the subcritical scattering we mean to 
discuss in the following sections.

\section{S-matrix in the subcritical case}
\label{scattering}

In the subcritical case, the scattering process is trickier than in the transcritical case. Taking into account the CTP makes even more difficult the analysis. 
Among the eight propagating modes (cf. Fig. \ref{fig-wh-sub} ), only modes $H,H',P,P',N,N'$ 
 are involved in the presence of the CTP \eqref{cctp} (which is, as  in sec. \ref{overb-landau} and in \cite{coutant-sub} the CTP nearest to the real axis with positive imaginary part), and one has to prescribe the behavior of each mode which is affected by the CTP. Instead, 
the spectator mode $V$ (and its counterpart $V'$ in the would-be white hole region) is not affected by the presence of the CPT, and then is involved only in the process  $V'\longrightarrow V$.\\

Our aim is to find the amplitudes relative to the following scattering process: 
\beq 
H \longrightarrow H'+P+N,
\label{pseudoH}
\eeq
which corresponds to the main process taken into account in subcritical studies, being also related to the experiments involving water. We stress that 
the process $V'\longrightarrow V$ represents a disconnected Feynman-like diagram in our picture, and this explains why $V,V'$ do not enter in \eqref{pseudoH}. 
We wish to find out $\alpha_\omega,\beta_\omega,\gamma_\omega$ such that
\beq
1=|\alpha_\omega|^2-|\beta_\omega|^2+|\gamma_\omega|^2, 
\label{amplitude}
\eeq
where 
\beqnl
|\alpha_\omega|^2:&=& \frac{|J_P|}{|J_H|}\\
|\beta_\omega|^2:&=& \frac{|J_N|}{|J_H|}\\
|\gamma_\omega|^2:&=& \frac{|J_{H'}|}{|J_H|},
\eeqnl
where the $J_S$, with $S=H,P,N,H'$ are the currents associated with the various modes in the asymptotic regions.\\

The main difficulty is how to prescribe a path,  i.e. a phase integral, in the complex plane such that modes $H,H',P,P',N,N'$ can actually feel the presence of the CTP.  
Our approach is a generalization to a fourth order ordinary differential equation of the approach considered in \cite{pokro} for a second order equation: therein, local connection formulas are exploited and also phase integrals. Another important point is that,  as to lateral connections,  we must keep into account a double 
contribution (cf. \cite{pokro}): on the one hand, we can set up lateral connections formulas {\sl locally} at the turning point, including the local effect of the Stokes curves, and then we must also consider nonlocal phase integral contributions \cite{heading}. One has to exploit WKB solutions as well as solutions for the near TP equation \eqref{four-z}. It is evident that, if the modes' domain is just the real line, there is no effect by the presence of a complex turning point and no mode conversion at all. In order to focus more precisely all the difficulties involved in the problem, we point out that is a further 
subtlety:  Stokes lines are present and are at the root of phenomena like the mode conversion phenomenon. For a fourth order equation and a sequence 
of complex turning points as in our case there would be no hope to keep under control the Stokes phenomenon and also the presence of new Stokes lines \cite{bnr}. One could numerically determine the network of Stokes lines 
associated with the sequence, but it would be hard to deduce any analytical expression.\\ 
Still, in the present case, we can take into account that Stokes lines 
are involved substantially with the dispersive modes and their dominant large momenta (cf. \cite{ni} and the Orr-Sommerfeld case). So we consider just the 
three Stokes lines associated with the fast dispersive modes, albeit even the exact geometry of the Stokes line might require a numerical calculation. 
In a neighborhood of the CTP, we get three Stokes lines: one vertical upward, and the other two at $120$ degrees from it. We also choose to 
put the branch cut arising from the CTP vertically upward as the vertical Stokes line. The effect of the Stokes lines is taken into account in the change of the basis B2-B1 occurring in the neighborhood of the CTP, by means of the local lateral connection formulas. Cf. subsection \ref{conne}.  
We can justify, {\sl a posteriori}, our approximation both by comparing with numerical simulations in literature (e.g. \cite{Parentani_numerical,parentani-sub}), 
and with the exact analysis in \cite{trevisan-exact}. 

\subsection{Phase integrals contributions}

\begin{figure}[hbtp!]
\centering
\includegraphics[scale=0.5]{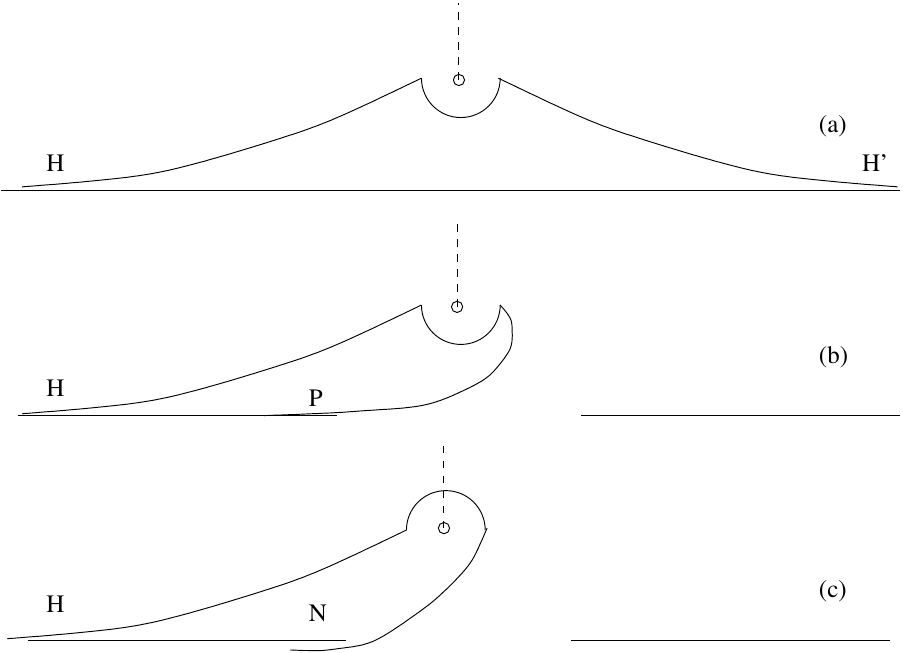} 
\caption{The ingoing mode is $H$, on the extreme left. In (a) we show the path we introduced for the calculation of \eqref{facth}, which is relative to the process $H \to H'$. 
In (b) we show the path we introduced for the calculation of \eqref{factp}, which is relative to the process $H \to P$. In (c) we show the path we introduced for the calculation of \eqref{factn}, which is relative to the process $H \to N$. The CTP, which corresponds to a pole, is indicated by a fat dot, and the branch cut appears as a vertical dashed line. In both cases (b) and (c), also the 
branch cut of ${\mathrm{arctanh} (z)}$, which plays a relevant role for $1/\epsilon$ contribution, is explicitly indicated.}
\label{sub-paths}
\end{figure}

Phase integrals involve the `semiclassical' action contributions 
\beq\label{phac}
S_i (x_{ctp},x):=\int_{x_{ctp}}^x ds\; k_i (s), \quad \quad i=H,H',P,N,
\eeq
where the $k_i$ are the momenta associated with each particle in the scattering process.  The case of fast momenta $k_P$ and $k_N$ include a leading term $\propto \frac{1}{\epsilon}$, and, with our velocity profile, we obtain  
\beq
\frac{1}{\epsilon} \int_{x_{ctp}}^x ds \sqrt{1-\frac{v^2 (s)}{c^2}}=\frac{1}{\epsilon}\frac{1}{\sqrt{6}} \int_{x_{ctp}}^x ds \sqrt{2-\tanh (s)},
\label{phase-int}
\eeq
which is proportional to the one (cf. equation \eqref{phase}) we have considered in the overbarrier scattering in the previous section. In such a way, by choosing 
$x\ll 0$ but finite, and approaching the cut of $\mathrm{arctanh}  (z)$ from below for $N$ and from above for $P$, cf. also Fig. \ref{sub-paths}, we find
\beq\label{sup-F}
F_\pm:=\exp \left (- \frac{\pi}{2 \sqrt{6} \epsilon}\right); 
\eeq
as a consequence, the same suppressing real factor multiplying the amplitudes of both the $P$ and the $N$ modes. This is, a posteriori, a nice ansatz. A strong suppression of $N$ with respect to $P$ could be also allowed, but is not our choice. We remark that 
our hypothesis of very weak $P,N$ production leads to the conclusion that 
\beq
\frac{|J_{H'}|}{|J_{H}|}\sim 1,
\label{transmitted}
\eeq
 which would be the naive result in the WKB approximation, and which is corrected by a small contribution due to the dispersive modes. In this approximation ansatz   
 we should obtain  
\beq
\frac{|J_{H'}|}{|J_{H}|}=1-O\left(\exp \left (-\frac{\pi}{\sqrt{6} \epsilon}\right)\right).
\label{true-transmitted}
\eeq
There are also contributions to be associated with a pole: they concern  the subdominant part 
\beq\label{expsub}
\exp (  i \frac{\omega}{c} \int_{x_{ctp}}^x ds \frac{v (s)}{c} \frac{1}{1-\frac{v^2 (s)}{c^2}}),
\eeq
with $x\ll 0$, of the dispersive modes $P,N$. Cf. also Fig. \ref{sub-paths}. We know, both on the ground of the well-known transcritical case, and on the ground of further analytical calculations 
(see e.g. \cite{trevisan-exact}), the $N$ mode (which remains an antiparticle mode also in the subcritical case) has a suppressed amplitude with respect to the 
$P$ mode (only in the limit for $\omega\to 0$  this suppression is heavily mitigated). In order to respect this indication, we choose for the $P$ mode 
a counterclockwise contribution below the turning point, i.e. the pole, which contributes a factor 
\beq\label{factp}
F_P=\exp (\frac{\pi \omega}{2\kappa})
\eeq
to its amplitude. For this kind of calculation, we refer to \cite{gamelin} (see in particular the 
so-called fractional residue theorem, p.209 in \cite{gamelin}). 
For the $N$ mode, we prescribe a clockwise contribution passing above the pole and, moreover, traversing the branch cut. This provides a 
factor 
\beq\label{factn}
F_N=\exp (-\frac{3\pi \omega}{2\kappa})
\eeq 
to its amplitude.\\  
There are further contributions to be potentially associated with a pole: a priori, concern both the Hawking ingoing mode, which, in agreement with its group velocity, is associated with the following contribution to the 
phase integral (cf. also Fig. \ref{sub-paths})
\beq
\varphi_H^{phase}=\exp \left( i \omega \int_{x}^{{x_{ctp}}} ds \frac{1}{c+v(s)} \right),
\eeq
with $x\ll 0$, and its outgoing `image' $H'$ in the right asymptotic region $x\gg 0$:
\beq
\varphi_{H'}^{phase}=\exp \left( i \omega \int_{{x_{ctp}}}^x ds \frac{1}{c+v(s)} \right). 
\eeq
Nevertheless, there is evidently a doubling of the contribution from the same pole in this branch of the scattering process, 
and this would not in the present case be compatible with \eqref{transmitted}, as we show in the following subsection. 
Then, we choose a counterclockwise semicircle passing below the turning point just only for $H$, and  
we obtain  the amplitude factor   
\beq\label{facth}
F_{H}=\exp (\frac{\pi \omega}{\kappa})
\eeq
for $H$. For $H'$ ve get $F_{H'}=1$ (no pole contribution). 
We display Fig. \ref{sub-paths}, where qualitative paths associated with the calculations of \eqref{facth}, \eqref{factp}, and  \eqref{factn} appear. Further information is found in the figure caption.

\subsection{Connection formulas and scattering amplitudes}
\label{conne}

We come back to the problem of how to determine the various amplitudes in the scattering process we are interested in. 
We know that, near the CTP, we can locally use central and lateral connection formulas. In the present case, at the 
turning point, compatibly with our choice of the phase integrals which must introduced in order to keep into account 
how we come back to the real axis, we get basis B2 on the left side 
and basis B1 on the right side of the CTP. 
Instead of using the functions $w_i (z)$ introduced in sec. \ref{NTP-corley}, we 
prefer to use directly the WKB functions they are connected to in the sense of central connection formulas. {It 
is to be stressed that, locally, the modes 
The relevant lateral local connection formula is 
\beq
\varphi_H  \mapsto  \varphi_{H'}+\left(1-\exp (-\frac{2 \pi \omega}{\kappa}) \right) \varphi_P,
\label{process}
\eeq
as it can be deduced by a careful inspection of the change of basis $B2\leftrightarrow B1$,  
whereas further informations about the ratio $|J_N|/|J_P|$, where $J$ is the conserved current in direction $x$, 
can be deduced directly by central connections and lateral phase integral connections.\\ 
We also recall that central connections are obtained in the so called linear region LR, where both the near turning point expansion 
and the WKB expansion coexist, by matching each expansion for each mode. In particular, one obtains 
\beq
\varphi_i^{WKB} \mapsto C_i \varphi_i^{WKB}, \quad \quad i=H,H',P,N,
\label{central-local}
\eeq
where $C_i$ are the central connections coefficients and the formula \eqref{central-local} holds true in LR. In the full scattering process at hand, 
we have to take into account the effect of the lateral connections, which are not only related to \eqref{process} if not just locally, indeed one must 
also take into account the further contributions arising from the phase integrals. By taking into account normalizations of the modes (see \cite{master,analog} for details),  
 we obtain \beqnl
|C_H|^2 &=& 4 \kappa c \exp (-\frac{2\pi \omega}{\kappa})\sinh \left( \frac{\pi \omega}{\kappa} \right),\\
|C_P|^2 &=& 2 \kappa c \exp (\frac{\pi \omega}{\kappa}),\\ 
|C_N|^2 &=& 2 \kappa c \exp (-\frac{\pi \omega}{\kappa}),\\ 
|C_{H'}|^2 &=& 4 \kappa c  \sinh \left( \frac{\pi \omega}{\kappa} \right).
\eeqnl
 
Let us first consider the ratio which, in the transcritical case, is considered as the signal of presence/absence of thermality: 
\beq
\frac{|J_N|}{|J_P|}=\frac{|C_N|^2 |F_N|^2}{|C_P|^2 |F_P|^2}= \exp (-\frac{6 \pi \omega}{\kappa}).
\label{t-ratio}
\eeq
Our choice for equally suppressed fast modes lets us obtain again a thermal-like behavior, with an effective temperature 
\beq
T_{eff}=\frac{T}{3},
\eeq
where $T$ is understood as the temperature for the transcritical case.  
We obtain a behavior that is qualitatively in agreement with the one discovered in \cite{trevisan-exact}, with a reduction of the effective temperature with respect to the 
transcritical case. We note that in \cite{trevisan-exact} a different physical case, involving both transcritical and subcritical cases in dielectrics, was studied. It seems that a somehow universal behavior, in this respect, is present also in the subcritical case.    
Let us then consider the process of pair creation/mode conversion: 
\begin{align}
\frac{|J_P|}{|J_H|}&=\exp \left (-\frac{\pi}{\sqrt{6} \epsilon}\right) \frac{\left(1-\exp (-\frac{2 \pi \omega}{\kappa}) \right)^2 |C_P|^2 |F_P|^2}{|C_H|^2 |F_H|^2} \cr
&=\exp \left (-\frac{\pi}{\sqrt{6} \epsilon}\right) \left(1-\exp (-\frac{2 \pi \omega}{\kappa})\right) \exp (\frac{\pi \omega}{\kappa}).
\end{align}
This formula has  the nice feature that, for $\omega \to 0$, 
 \beq
 \frac{|J_P|}{|J_H|} \sim \exp \left (-\frac{\pi}{\sqrt{6} \epsilon}\right) \frac{2 \pi \omega}{\kappa}  \propto \omega,
 \eeq
 as expected, and then, thanks to \eqref{t-ratio}, 
\beq
 \frac{|J_N|}{|J_H|}= \frac{|J_P|}{|J_H|} \exp (-\frac{6 \pi \omega}{\kappa})
\propto \omega,
 \eeq
in the same limit. This means that the thermal spectrum behavior which occurs in a transcritical (white hole) situation is not present, in agreement also with the aforementioned numerical 
calculations in literature.  One may wonder if the factor  $\exp (\frac{\pi \omega}{\kappa})$ might be problematic. We can remark that, in our approximation, 
$\omega$ is considered to be $O(1)$, and then the first exponential factor $\exp \left (-\frac{\pi}{\sqrt{6} \epsilon}\right)$ largely dominates it. Still, we stress also that in the scattering with dispersion, there exists also 
maximal frequency $\omega_{max}$ such that there is no Hawking process for $\omega>\omega_{max}$ \cite{corley-jacobson,macher-d,macher-bec}. Such a $\omega_{max}$ is proportional to $\frac{1}{\epsilon}$, and tends to infinity as $\epsilon \to 0$ \cite{macher-d,macher-bec,cpf}. We cannot approach analytically the maximal frequency, as it would mean to revise the WKB approximation in a regime analytically very hard to manage.\\
As to the transmitted flux ratio,  the ratio  
\eqref{transmitted} just satisfies \eqref{true-transmitted} without the presence of suppression factors $\propto \exp (-\alpha \omega)$, with $\alpha>0$, which would be 
hard to believe, at least in the region where $\omega=O(1)$.\\
It is also interesting to stress that, by taking into account the small correction appearing in \label{true-transmitted} and improving consequently our transmitted flux, qualitatively we find a behavior in agreement with the one in \cite{parentani-sub}: 
\beqnl\label{effective}
|\gamma_\omega|^2&=& \frac{|J_{H'}|}{|J_H|}=1-(|\alpha_\omega|^2-|\beta_\omega|^2)\cr
&=& 1-\exp \left (-\frac{\pi}{\sqrt{6} \epsilon}\right) \left(1-\exp (-\frac{2 \pi \omega}{\kappa})\right) \exp (\frac{\pi \omega}{\kappa})\left(1-\exp (-\frac{6 \pi \omega}{\kappa})\right).
\eeqnl
Indeed, we get a reduction of the transmission probability $H\to H'$ which increases as $\omega$ becomes greater.\\ 
 
Let us discuss again our results in light of the existent literature on this topic.
In the subcritical case, for which we refer also to the 
discussion in \cite{barcelo} (see pp. 126-128), there is still a pair creation process. Let us provide an intuitive kinematic interpretation in the real domain for the pair creation occurring in this horizonless configuration, which physically illuminates our previous complex turning point (CTP) analysis. For a subluminal dispersion relation, such as the one characterising the Corley model, high-wavenumber dispersive modes propagate with an effective phase (or group) velocity that (for $\epsilon\neq 0$) is smaller than the low-frequency sound speed $c$. Consequently, even if the background flow is globally subcritical (i.e., $\vert v(x) \vert < c$ everywhere, meaning that no standard acoustic horizon is formed for low frequencies), it can locally exceed the effective velocity of these short-wavelength modes. This leads to the formation of a frequency-dependent ``effective horizon'' (or effective ergoregion) tailored for these specific high-frequency quanta, triggering a process similar to Cherenkov radiation or localised anomalous mode conversion \cite{barcelo}. This physical mechanism shares deep physical roots with quantum vacuum radiation induced by refractive index perturbations in dielectric media where geometric horizons are strictly absent \cite{belgiorno-prl}. It explains the puzzling occurrence of excitations in subcritical flows, a feature captured by the `complex horizon' appearing in \cite{coutant-subcritical}, which is at a zero order step frequency independent and then is corrected perturbatively and becomes frequency dependent. This same feature is shared also 
by alternative semiclassical frameworks like the tunnelling method \cite{delporro}. In a semiclassical analysis involving complex paths and 
approximant trajectories, 
an effective horizon appears therein, which is still blocking and frequency dependent. 
It appears at high frequencies (below the maximal frequency). 
Also within our Orr-Sommerfeld-like fourth-order framework, this real-space effective horizon seems to be mathematically translated into the existence of complex turning points, which act as subtle regulators for the overbarrier-like mode conversion. In our framework, our CTP is not $\omega$-dependent. Still, we can analytically calculate the amplitudes associated with the process, 
and find that, despite the absence of a real TP, and of a real horizon, there is a pair-production process, and \eqref{effective} shows that in our self-consistent approach we get a reduction of the 
transmission ratio for the Hawking particle which is consistent with the existence of an effective horizon which is frequency dependent, as the transmission decreases as the frequency increases. 
In this sense, our calculation is just consistent and compatible with both numerical calculations (see e.g. \cite{parentani-sub}) and with analytical ones \cite{coutant-subcritical,delporro}. Moreover, 
it is consistent also with perturbative results \cite{trevisan} and with exact results \cite{trevisan-exact}, where the decreasing of the transmission coefficient for $\omega$ increasing (but always below $\omega_{max}$) is a posteriori associated with the presence of an effective frequency dependent horizon, as described above.   
To sum up,  in our view, the aforementioned methods, including ours one, represent  different approaches to the same problem, clarifying the picture concerning the subcritical case starting from different approximations.

\section{Further critical considerations}
\label{critical}

As well known in  the literature, checking and measuring only the ratio \eqref{t-ratio} is not sufficient for being in the presence of the Hawking effect, and we have found a heavily reduced 
particle creation. Still, this is by no means the end of the story. If we restore the factor $\gamma$ in the profile \eqref{profv}, we obtain 
\beq
x_{ctp}= \frac{x_0+i \frac{\pi}{2}}{\gamma}.
\label{tanh-ctp}
\eeq
It is very interesting to notice that, the larger the steepness $\gamma$, the smaller the imaginary part becomes, which means that 
the steepness $\gamma$ of the step-like profile is a very important element in getting a complex turning point as near as possible to the real axis, so qualitatively 
corroborating the occurrence of a still efficient pair-production in the subcritical regime. Notice also that the derivative of \eqref{profv} is 
\beq
v'(x)= \frac {c^2}v  \frac {\gamma}{12\cosh^2 (\gamma x)},
\label{hyptang}
\eeq
so that a large $\gamma$ enhances the value of the surface gravity $\kappa=v'(x_{ctp})= \frac {c}{4} \gamma$ with respect to the case where $\gamma=1$. We could also generalize the profile \eqref{profv} by choosing 
e.g. 
\beq\label{profab}
v(x):=-c \left(d+b\tanh (\gamma x) \right), 
\eeq
with  $d,b>0$, we find 
\beq
x_{ctp}= \frac{\tanh^{-1} (\frac{1-d}{b})+i \pi n}{\gamma}, 
\eeq
with integer $n\in {\mathbb{Z}}$,  in such a way  that one is in the subcritical case for $1-d>b$, 
and a complex turning point appears, with a real part which is as more near to 0 as bigger and positive is the ratio $(1-d)/b$, and an imaginary part which can be chosen to be  always $\pi/2 \gamma$, and, for $\gamma\gg 1$, one finds $0<\pi/2 \gamma\ll 1$, i.e., even the imaginary part may be very near the real axis. As a consequence, an enhancement of pair-production even in the subcritical case can be expected in such a situation. In particular, the suppression factor \eqref{sup-F} is replaced by 
\beq
F_s:=\exp \left (-\frac{\pi}{2 \sqrt{6} \gamma \epsilon}\right).  
\label{supf}
\eeq
In a realistic situation of weak (but non-zero) dispersion, represented by $\epsilon$, and with a very strong $\gamma\gg 1$, which amounts to a very strong 
surface gravity, that suppression factor could be reasonably be reduced very near 1, and then would allows a possibly good rate of pair creation, if 
$\gamma \epsilon$ is not a very small number. This might be decisive in order to understand results obtained in dielectric media, but also in water experiments. 
It appears that, in subcritical regimes, a big $\gamma$ might play a fundamental role.\\

A final consideration is relative to the critical case discovered in \cite{trevisan-exact}. 
Let us consider the following generalized profile:
\beq\label{gprof}
v(x) = -c \sqrt{\frac{2}{3}+b \tanh x},
\eeq
where $b$ is such that the subcritical case occurs for $0<b<\frac{1}{3}$ and the transcritical one for $b>\frac{1}{3}$. As we see, there is a critical value 
\beq
b_{crit}=\frac{1}{3}
\eeq
which displays a very peculiar behavior.   Indeed, at this critical value the turning point, which is otherwise present, is absent. There is no possible solution, 
even in the complex plane. This means that any strategy involving real/complex turning points is doomed to failure. The standard semiclassical approximation 
still holds true everywhere, and the scattering process becomes simply $H\to H'$ with probability 1. Moreover, there is no more pair production. 
The critical case, which, remarkably, could be reached both from the subcritical case and from the transcritical one, shares in this sense some weak resemblance with the extremal 
black hole case, because naively there is a zero temperature, which is compatible with the naive limit of \eqref{hyptang} for 
$x\to +\infty$. But we don't claim at all that this is the analog extremal case,  extremal black holes in the astrophysical regime are characterized by a degenerate horizon, which is missing herein, and, moreover, they are still affected by a phenomenon of particle emission due to the loss of charge and/or the loss of angular momentum. 
 Anyway, what is relevant for us is that, a posteriori, in our framework the existence of $b_{crit}$ justifies the result reached 
in \cite{trevisan-exact}. We notice that, for the more general profile \eqref{profab} the critical value would be  $b_{crit}=1-d$.

\section{Conclusions}\label{concl}

In the present paper, we have analytically calculated the main amplitude involving pair creation in the white hole subcritical case. We point out the difference with the 
transcritical case: in the subcritical process, there is no particle playing the role of Hawking partner, as $H'$ in the would-be white hole region 
is just an outgoing particle and not an antiparticle traveling towards the horizon. Furthermore, in the 
transcritical case, the growing mode $G$ must be set to zero for obvious physical reasons,  and the 
decaying mode $D$, which instead plays a fundamental role in the Hawking process of production of the Hawking particle $H$, 
even if it does not contribute to the current conservation, of course. $G$ and $D$ are replaced, in the subcritical case, by the 
modes $P'$ and $N'$, both ingoing from  $x=\infty$. As to the process $V\to V'$, it is trivial in both cases.\\ 

In the subcritical case, the process physically most interesting is the one involving $H$ as ingoing particle, 
i.e. the would-be Hawking particle. In order to deal also with the subcritical case, we must involve complex turning points, in a similar way to 
what happens in standard quantum mechanics in the so-called overbarrier scattering, where the reflected mode can be recovered only by 
introducing CTPs. This is true also in the present case, where a naive WKB approach would not be able to give rise to mode conversion. We exploit 
central connections and lateral connections again, the latter ones being completed, with respect to the local ones, by important phase integral contributions. Cf. 
also \cite{pokro}. 
Also in \cite{coutant-subcritical} CTP are recovered within a different approach involving the so-called Bremmer approximation, suitably generalized to 
$n-th$ order ordinary differential equations, and therein a concept of complex horizon is also introduced.  Also, in \cite{delporro} there appears an effective horizon which is associated to the semiclassical tunneling framework also for the subcritical case, as we recalled in the concluding 
part of sec. \ref{scattering}.    
Herein, our different approach again 
assigns a fundamental role to CTP, still we prefer to avoid to refer to it as a complex horizon, but simply as an effective horizon, and allows to treat the subcritical case with same analytical tools as in the transcritical one, albeit in a much trickier situation and with the need to appeal to a number of hypotheses whose effectiveness can be judged only a posteriori, by comparison with numerical data and further analytical informations by other studies \cite{trevisan-exact}.   
Results for the probability of creation of the dispersive  $N,$ $P$ modes have nice analytical features, as remarked above. 
Furthermore, 
there is also a nice effective thermal character for the ratio \eqref{t-ratio}, with a reduced temperature which is corroborated by results in \cite{trevisan-exact}. 
The particle production 
process can be not so much suppressed with respect to the main process $H\to H'$ in presence of a sufficiently big $\gamma$. Last, we are able also to justify the presence of a 
critical value of the parameter multiplying $\tanh(\gamma x)$ in the expression \eqref{gprof}. It is characterized by the absence of turning points, and is involved with the absence of pair creation.\\ 

One might wonder if a subcritical case could exists also for the superluminal case, where dispersive propagating modes 
appear only in the black hole region where $|v|>|c|$. Given that $V\mapsto V'$, in our framework of weak dispersion, plays the role of a disconnected 
line in a Feynman diagram, what happens if $|v|<|c|$ everywhere on the real line, as discussed for the subluminal case, is that no contribution arises from dispersive modes, 
for what concerns fluxes at infinity, so that one would obtain trivially $|J_H|=|J_{H'}|$. A condition analogous to the subluminal subcritical case, with 
dispersive propagating modes could be found also in the external region, could  
be reached only for $|v|>|c|$ everywhere. 
But, actually, one would implement a white hole condition everywhere (analogously for a black hole), which physically would be neither meaningful nor feasible.\\

In conclusion,  
we still consider the subcritical case as 
a process with maintain an interesting but weak relationship with the Hawking effect. It is true that, if one observes only what happens in the external region, the 
same four modes appear: $H,P,N,V$, with only $H$ as ingoing mode. Still, in the would-be white hole region, changes are dramatic, as the only mode 
which is the same as in the transcritical case is the spectator mode $V'$. And also the phenomenon itself changes, as discussed above, and as 
 the  the spectrum is not Planckian, even if the ratio \eqref{t-ratio} can be associated with a temperature, as usual. The subcritical regime is also separated from the transcritical one by a critical regime where no particles are produced.

\appendix
\section{The change of basis according to \cite{ni,ni-global}}
\label{ba-nishi}

As discussed in detail in \cite{analog}, for the subluminal case we can provide the following lateral connection formulas, representing the relations between 
the bases, in perfect agreement with the discussion of the superluminal case appearing in \cite{ni,ni-global} (for a further discussion of the superluminal case we refer to \cite{analog}):
\begin{align}
[1,w_4(z),w_1(z),w_2(z)] &=[1,w_6(z),w_3(z),w_1(z)] P_3, \label{pitre}\\
[1,w_5(z),w_2(z),w_3(z)] &=[1,w_4(z),w_1(z),w_2(z)] P_2, \label{pidue}\\
[1,w_6(z),w_3(z),w_1(z)] &=[1,w_5(z),w_2(z),w_3(z)] P_1, \label{piuno}
\end{align}
where the matrices $P_i,i=1,2,3$ connect the bases above, and, in the following, will be explicitly written at the leading order in the series in $\epsilon$. It is not difficult to see that, in considering the relation between B1 and B2,  it holds \cite{analog}
\beq\label{pp3}
P_3=
\left(
\begin{array}{cccc}
1 & 0 & 0 & 0\cr
0 & 1 & 0 & 1\cr
0 & 0 & 0 & -1\cr
0 & 1-\psi^{-3 \lambda} & 1 & -\psi^{-3 \lambda}
\end{array}
\right).
\eeq
Analogously, as far as the relation between B3 and B1 is concerned, one finds
\beq\label{pp2}
P_2=
\left(
\begin{array}{cccc}
1 & 0 & 0 & 0\cr
0 & \psi^{-3 \lambda} & 0 & 1\cr
0 & 0 & 0 & -1\cr
0 & 1-\psi^{-3 \lambda} & 1 & -1
\end{array}
\right).
\eeq
As to the relation between B2 and B3, analogously, one obtains \cite{analog}
\beq\label{pp1}
P_1=
\left(
\begin{array}{cccc}
1 & 0 & 0 & 0\cr
0 & 1 & 0 & \psi^{3 \lambda}\cr
0 & 0 & 0 & -\psi^{3 \lambda}\cr
0 & 1-\psi^{-3 \lambda} & 1 & -1
\end{array}
\right).
\eeq

\section{Bases and lateral connection formulas according to \cite{langer1955}}
\label{ba-langer}

In \cite{langer1955}, a long and detailed discussion of the solutions of the equation 
\beq\label{langer-eq}
v'''+\lambda^2 z v'+3 \mu \lambda^2 v=0
\eeq
appears. It must be stressed that, in this Appendix, we maintain the same notation as in the original 
paper \cite{langer1955}, and then one might be confused as $z$ is not the same variable we used in the main text 
and $\lambda$ does not keep the same meaning as well, as it is a large complex parameter in \cite{langer1955}. There 
exists an evident correspondence between equation \eqref{four-z} and equation \eqref{langer-eq}, which is 
obtained as follows: in \eqref{langer-eq} let us set $z=\lambda^{-2/3} x$. Cf. also \cite{vallee}. Then we obtain 
\beq
\frac{d^3 v}{dx^3}+ x\frac{d v}{dx}+3 \mu  v=0.
\eeq
We remark that solutions in \cite{langer1955} are expressed in terms 
of the variable $\xi=(i \lambda z^{3/2})/3$. We shall translate all results in terms of the variable $z$ used in the main text 
(which amounts to set $\xi=(i  z^{3/2})/3$).\\
The identification 
\beq
v=\frac{d\zeta}{dx},
\eeq
and the trivial substitutions $x\mapsto z$ and $3 \mu\mapsto \lambda-1$ (where henceforth $\lambda$ is meant to be the parameter defined in \eqref{lambda-n},  cf. also \cite{ni-global}) make evident the one-to-one 
mapping of the aforementioned equations and of the relative solutions.\\
For each integer $q$, Langer introduces three different functions, for which he also provides the asymptotic behavior, also truncated at the leading order and with the parameters of the Corley model:
\beq
v_{q,0}(z) \sim \xi^{-2\mu} \sum_{n=0}^\infty \frac{\kappa_n}{(3 i \xi)^n}\sim \left(\frac{i}{3}\right)^{\frac{2 i\omega}{3 \kappa}} z^ \frac{i\omega}{\kappa}+\cdots
\eeq
which holds for 
\beq
\frac{2}{3} (q-2) \pi <\arg (z) < \frac{2}{3} q \pi;
\eeq
\beq
v_{q,1}(z)\sim \xi^{\mu-1/2} e^ { 2 (-)^q \xi} \sum_{n=0}^\infty \frac{\nu_n}{(-3 i (-)^q \xi)^n} \sim  \left(\frac{i}{3}\right)^{-\frac{i\omega}{3 \kappa}-\frac{1}{2}} z^{-\frac{i\omega}{2\kappa}-\frac{3}{4}} e^ { i\frac{2}{3} (-)^q z^\frac{3}{2}}+\cdots  ,
\eeq
which holds for 
\beq
\frac{2}{3} (q-3) \pi <\arg (z) < \frac{2}{3} q \pi;
\eeq
\beq
v_{q,2}(z)\sim \xi^{\mu-1/2} e^ {- 2 (-)^q \xi} \sum_{n=0}^\infty \frac{\nu_n}{(3 i (-)^q \xi)^n}\sim  \left(\frac{i}{3}\right)^{-\frac{i\omega}{3 \kappa}-\frac{1}{2}} z^{-\frac{i\omega}{2\kappa}-\frac{3}{4}} e^ {- i\frac{2}{3} (-)^q z^\frac{3}{2}}+\cdots ,
\eeq
which holds for 
\beq
\frac{2}{3} (q-2) \pi <\arg (z) < \frac{2}{3} (q+1) \pi.
\eeq
Then we can reconstruct the three bases B1,B2,B3 inherited by Nishimoto \cite{ni,ni-global}. We stress that, in order to recover the ordered bases which are used in the change of basis rules described in the previous Appendix \ref{ba-nishi}, we must change order to the exponential contributions in the bases one would naturally write in terms of Langer's analysis.\\
B1 corresponds to $\left\{ 1, v_{1,0}(z), v_{1,2}(z), v_{1,1}(z)\right\}$, for which, indeed, we have 
$-\frac{2}{3} \pi <\arg (z) < \frac{2}{3} \pi$. 
Notice that just $v_{1,0}(z)$ has exactly this argument 
(we have $\arg (z)\in (-\frac{4}{3} \pi,\frac{2}{3} \pi)$ for $v_{1,1}(z)$ and $\arg (z)\in (-\frac{2}{3} \pi,\frac{4}{3} \pi)$
for $v_{1,2}(z)$).\\
B2 corresponds to $\left\{ 1, v_{2,0}(z), v_{2,2}(z), v_{2,1}(z)\right\}$, for which, indeed, we have 
$0 \ <\arg (z) < \frac{4}{3} \pi$. Notice that just $v_{2,0}(z)$ has exactly this argument 
(we have $\arg (z)\in (-\frac{2}{3} \pi,\frac{4}{3} \pi)$ for $v_{2,1}(z)$ and $\arg (z)\in (0 ,2 \pi)$
for $v_{2,2}(z)$).\\
B3 corresponds to $\left\{ 1, v_{3,0}(z), v_{3,2}(z), v_{3,1}(z)\right\}$, for which, indeed, we have 
$\frac{2}{3} \pi  <\arg (z) < 2 \pi$. Notice that just $v_{3,0}(z)$ has exactly this argument 
(we have $\arg (z)\in (0\pi,2 \pi)$ for $v_{3,1}(z)$ and $\arg (z)\in (\frac{2}{3} \pi, \frac{8}{3} \pi)$
for $v_{3,2}(z)$).\\
We stress that there is a correspondence in the asymptotic behavior in $z$, but not in the coefficients, which still need to be suitably adjusted in order to get a full correspondence.\\
Langer does not provide a direct formula for a basis change, just because he provides a more general 
formula relating a generic element $v_{q,i}(z)$ to the three elements of a basis of \eqref{langer-eq}:
\beq\label{langer-lateral}
v_{q,i}(z)=\delta^{(q,i)}_{r,0} v_{r,0}(z)+\delta^{(q,i)}_{r,1} v_{r,1}(z)+\delta^{(q,i)}_{r,2} v_{r,2}(z), \quad \quad i=0,1,2,
\eeq
where the coefficients $\delta^{(q,i)}_{r,0},\delta^{(q,i)}_{r,1},\delta^{(q,i)}_{r,2}$ are given in \cite{langer1955}, equations 
(9.5),(9.6) and (9.7) therein. In particular, we refer to \cite{langer1955}, p. 537 (but not only) for full details. It is a bit tedious but possible to show that the change of basis 
indicated in Appendix \ref{ba-nishi} correspond to the ones obtainable by the bases indicated herein and inherited by 
Langer \cite{langer1955}.\\
In concluding this section, we  also recall that, as discussed also in \cite{langer1955}, solutions at hand can also be related to the 
generalized hypergeometric functions. As indicated in \cite{analog}, 
the general exact solution of \eqref{four-z}, can be expressed in terms of the generalized hypergeometric function $\,_1F_2 $ as follows: 
\begin{align}
\label{exact-near}
w(z)&= C_1+C_2 z \,_1F_2 \Big(\frac{\lambda}{3};\frac{2}{3},\frac{4}{3};-\frac{z^3}{9}\Big)+C_3 z^2 \,_1F_2 \Big(\frac{1+\lambda}{3};\frac{4}{3},\frac{5}{3};-\frac{z^3}{9} \Big)\cr
&+ C_4 \,_1F_2 \Big(\frac{-1+\lambda}{3};\frac{1}{3},\frac{2}{3};-\frac{z^3}{9}\Big).
\end{align}
But the drawback of this solution is that the physical modes we are interested in are tricky linear combinations of the hypergeometric function $\,_1F_2 $ 
indicated above.



\end{document}